\documentclass[aps,prd,onecolumn,superscriptaddress,showpacs,amsmath,amssymb,amsthm,nofootinbib, preprintnumbers]{revtex4-2}
\usepackage{amsmath}
\usepackage{graphicx}
\usepackage{epstopdf}
\usepackage{float}
\usepackage{hyperref}
\usepackage{color}
\usepackage[T1]{fontenc}
\usepackage[utf8]{inputenc}
\usepackage[toc,page]{appendix}
\usepackage[usenames,dvipsnames]{xcolor}
\usepackage[normalem]{ulem}
\usepackage{lipsum, babel}
\usepackage[justification=justified,format=plain]{caption}

\usepackage{subcaption}
\usepackage[utf8]{inputenc}
\usepackage{url}
\usepackage{newunicodechar}
\newunicodechar{−}{-}
\usepackage{xcolor}
\usepackage{physics}

\usepackage{hyperref}

\usepackage{graphicx}
\usepackage{dcolumn}
\usepackage{bm}

\newcommand{\be}{\begin{equation}}             
\newcommand{\ee}{\end{equation}}               
\newcommand{\ba}{\begin{eqnarray}}
\newcommand{\ea}{\end{eqnarray}}

\begin{document}

\title{A Cardy-like expression for {charged} rotating solitons and black holes}
\author{Moises Bravo-Gaete}
\email{mbravo@ucm.cl, moisesbravog@gmail.com}
\affiliation{Departamento de Matem\'atica, F\'isica y Estad\'istica, Facultad de Ciencias
B\'asicas, Universidad Cat\'olica del Maule, Casilla 617, Talca, Chile.}

\author{Fabiano F. Santos}
\email{fabiano.ffs23@gmail.com}
\affiliation{School of Physics, Damghan University, Damghan, 3671641167, Iran.}
\affiliation{Departamento de Física, Universidade Federal do Maranhão, São Luís, 65080-805, Brazil.}
\affiliation{Centro de Ciências Exatas, Naturais e Tecnológicas, UEMASUL, 65901-480, Imperatriz, MA, Brazil.}

\author{Xiangdong Zhang}
\email{scxdzhang@scut.edu.cn}
\affiliation{School of Physics and Optoelectronics, South China University of Technology, Guangzhou 510641, China.}

\date{\today}          
\begin{abstract}
This paper aims to propose a Cardy-like formula characterized by the mass, charge, and angular components of the black hole, along with their corresponding solitonic configuration, obtained through a double Wick rotation.  The expression also incorporates the dynamical exponent and effective spatial dimensionality as key elements. 
To validate the proposal, we first present a new concrete example in which recovering the semiclassical entropy requires the soliton to possess thermodynamic quantities beyond its mass. Additionally, we show more examples derived from static black hole solutions, employing a Lorentz boost to calculate their thermodynamic parameters. Finally, we include a case of a rotating configuration where the Lorentz boost is not required.
\end{abstract}

\maketitle
\section{Introduction}

Black holes (BHs) have evolved into objects of profound interest, as evidenced by empirical discoveries such as gravitational waves \cite{TheLIGOScientific:2017qsa,Monitor:2017mdv} and the imaging of {BHs} in the galaxies Messier 87 \cite{Akiyama:2019cqa,Akiyama:2019brx,Akiyama:2019sww,Akiyama:2019bqs, Akiyama:2019fyp,Akiyama:2019eap} as well as Sagittarius A* \cite{EventHorizonTelescope:2022apq,EventHorizonTelescope:2022wok,EventHorizonTelescope:2022exc,EventHorizonTelescope:2022urf,EventHorizonTelescope:2022xqj}. Particularly in the 1970s, often referred to as the golden age of BH research, significant contributions from Hawking, Carter, and Bardeen \cite{Bardeen:1973gs,Hawking:1974sw,Bekenstein:1973ur} illuminated the thermodynamic behaviors of these objects. The analogies established between area and entropy, as well as surface gravity and temperature, laid the groundwork for the thermodynamic laws of BHs, which parallel those of classical thermodynamics. Moreover, the second law of thermodynamics enables us to analyze the stability of these systems by examining their responses to fluctuations in quantities, such as energy and temperature, around equilibrium.

These ideas have proven very useful for understanding key concepts related to the Anti de Sitter (AdS)/Conformal Field Theory (CFT) correspondence \cite{Maldacena:1997re,Gubser:1998bc,Witten:1998qj,Klebanov:2009zz,Plefka:2005bk}. For instance, in two-dimensional CFTs \cite{Belavin:1984vu}, Cardy et al. \cite{Cardy:1986ie,Bloete:1986qm} derived an expression for the asymptotic density of states, leading to an expression for the entropy of a CFT. On the gravity side, Brown and Henneaux \cite{Brown:1986nw} noted that the asymptotic symmetries of a three-dimensional Einstein-Hilbert theory with a cosmological constant correspond to two copies of the Virasoro algebra; this can also be described by a two-dimensional CFT with a central charge. Building on this, Strominger \cite{Strominger:1997eq} confirmed that with this central charge and a fundamental state identified with the three-dimensional AdS spacetime, the Cardy entropy aligns perfectly with the Bekenstein-Hawking entropy of the Ba\~nados-Teitelboim-Zanelli (BTZ) BH \cite{Banados:1992wn}. 

However, as shown in \cite{Correa:2010hf,Correa:2011dt}, where hairy configurations in three dimensions were studied, there are concrete examples where {a gravitational soliton can describe the fundamental state. For the sake of completeness, one can study phase transitions between the AdS planar BH and the AdS soliton, obtained via a double analytic continuation, and which acts as the thermal ground state \cite{Surya:2001vj}. According to the interpretation put forward in \cite{Witten:1998qj,Witten:1998zw}, the deconfined phase is represented by a large, cold planar BH, while the confined phase corresponds to the small, hot AdS soliton \cite{Surya:2001vj}.}

Additionally, AdS/CFT duality extensions have facilitated the study of gravitational dual theories in {$d$}--dimensions, particularly those with non-standard asymptotic behavior. One such extension is known as the hyperscaling violation metric, which is defined by \cite{Charmousis:2010zz}
\begin{align}\label{eq:Hyperscaling}
ds^2_{H}={\left( \frac{1}{r} \right)^{\frac{2\theta}{d-2}}} \left(-{r^{2z}} \,dt^{2}+\frac{dr^{2}}{r^{2}}+{r^{2}} {d}\varphi^2+{r^2 \sum_{i=1}^{d-3} dx_i^2}\right).
\end{align}
Here, $-\infty<t<+\infty$, $r\geq 0$, $0\leq \varphi<2\pi$, {$0\leq x_i<\sigma_i$,} while that $z$ and $\theta$ are known as the dynamical exponent and hyperscaling violation exponent, respectively, where for $z=1$ and $\theta=0$ the AdS metric is recovered. In addition, we note that the space-time (\ref{eq:Hyperscaling}) enjoys a conformal transformation $ds^2_{H}\rightarrow {\tilde{\lambda}^{\frac{2 \theta}{d-2}}} ds^2_{H} $ with  $
t\to\tilde{\lambda}^z\,t, \,\,  \varphi \to\tilde{\lambda}
\varphi$ and $
r\to\tilde{\lambda}^{-1} r,
$  being $\tilde{\lambda}$ a constant. To study the correspondence to a finite temperature, static BHs must be introduced,  where their asymptotic behavior reproduces the metric (\ref{eq:Hyperscaling}). These configurations are supported with many alternative gravitational theories, characterized by the introduction of matter sources \cite{Shaghoulian:2015dwa} or higher-derivative correction terms \cite{Bravo-Gaete:2015wua}. This kind of geometry has emerged as a significant extension of the gauge/gravity duality, providing gravitational backgrounds that can holographically describe non-relativistic and scale-anomalous quantum field theories. For example, in \cite{Dong:2012se}, the holographic entanglement entropy uncovers the presence of novel phases that violate the conventional area law. In these regimes, the entropy exhibits a scaling behavior that interpolates between that of systems with hidden Fermi surfaces and those characterized by extensive entanglement, signaling rich underlying quantum correlations. 

From the perspective of the dual CFT, an anisotropic Cardy formula was proposed in \cite{Gonzalez:2011nz}  by using the isomorphism between two-dimensional Lifshitz algebras with dynamical exponents $z$ and $z^{-1}$ respectively, assuming the soliton mass as the ground state separated by a gap from the BH spectrum. Building on this, in \cite{BravoGaete:2017dso,Shaghoulian:2015lcn}, the authors proposed an extension of this generalized Cardy expression in arbitrary dimensions based on the vacuum energy with a gravitational soliton (see Eqs. (2.4)–(2.5) in Ref. \cite{BravoGaete:2017dso} and Eq. (28) in Ref. \cite{Shaghoulian:2015lcn}). Nevertheless, as we will see in the following lines, we can find an example in the literature where, to recover the semiclassical entropy, the soliton must possess more thermodynamic parameters.

Motivated by the preceding discussion, we now consider the generalized Smarr relation for charged and rotating BHs \cite{Smarr,Brenna:2015pqa} \footnote{It is possible to consider a more general case such as: ${\cal{M}}=\left(\frac{d_{\tiny{\mbox{eff}}}}{z+d_{\tiny{\mbox{eff}}}}\right)T {\cal S}+\alpha_e \Phi_e {\cal Q}_e+\eta \Omega {\cal{J}}$ (see for example Ref. \cite{Bravo-Gaete:2025lgs}). Nevertheless, in all the examples below, we will consider the constant $\eta$ to be unity.}:
\begin{equation}\label{smarrrotating}
{\cal{M}}=\left(\frac{d_{\tiny{\mbox{eff}}}}{z+d_{\tiny{\mbox{eff}}}}\right)T {\cal{S}}+\alpha_e \Phi_e {{\cal
Q}}_e+\Omega {\cal{J}},
\end{equation}
where ${\cal{M}}$ denotes the mass of the BH, $\Phi_e$ and ${\cal{Q}}_e$ correspond to the electric potential and electric charge, respectively, and $\alpha_e$ is a theory-dependent constant associated with the electromagnetic sector. The quantities $\Omega$ and ${\cal{J}}$ denote the angular velocity and angular momentum. As before, $z$ is the dynamical exponent and $d_{\tiny{\mbox{eff}}}$ is the effective spatial dimensionality, which reflects the scale of the entropy ${\cal{S}}$ with respect to the temperature $T$ in the following way \cite{Gouteraux:2011ce,Sachdev:2012dq,Charmousis:2012dw}
\begin{equation}\label{eq:S-T}
{\cal{S}}\propto T^{\frac{d_{\mbox{\tiny{eff}}}}{z}}.
\end{equation}
Note that for {$d_{\mbox{\tiny{eff}}}=d - 2$} we have the anisotropic case \cite{Gonzalez:2011nz,BravoGaete:2017dso,Shaghoulian:2015lcn} while the isotropic situation is recovered when {$d_{\mbox{\tiny{eff}}}=d-2$, $z=1$} and $\theta=0$. Nevertheless, for hyperscaling violation BHs according to the line element (\ref{eq:Hyperscaling}), the effective spatial dimensionality depends explicitly on the gravity theory considered (see, for example, Refs.  \cite{Bravo-Gaete:2015wua,Bravo-Gaete:2015iwa,BravoGaete:2017dso}).

From the Smarr formula \eqref{smarrrotating} and Eq. (\ref{eq:S-T}), it follows \textcolor{black}{that
$$T {\cal{S}} \propto {\cal{S}}^{\frac{z+d_{\mbox{\tiny{eff}}}}{d_{\mbox{\tiny{eff}}}}} \propto \left({\cal{M}}-\alpha_e \Phi_e {{\cal
Q}}_e-\Omega {\cal{J}}\right),$$
and the entropy can also be expressed as}
$${\cal{S}} \propto \left({\cal{M}}-\alpha_{e}\, \Phi_e {\cal Q}_e-\Omega {\cal{J}}\right)
^{\frac{d_{\mbox{\tiny{eff}}}}{z+d_{\mbox{\tiny{eff}}}}}.$$
In this work, we propose that the proportionality constant connecting these thermodynamic quantities is determined by thermodynamic parameters characterizing the corresponding solitonic configuration. This leads to a Cardy-like expression for the entropy, given by
\textcolor{black}{\begin{eqnarray}\label{Cardyrotatinghvm}
{\cal{S}}_{C}&=&\frac{2 \pi(z+d_{\mbox{\tiny{eff}}})
}{d_{\mbox{\tiny{eff}}}}
\left[-\frac{d_{\mbox{\tiny{eff}}}}{z} \left({\cal{M}}_{\mbox{\tiny{sol}}}+
\Omega_{\mbox{\tiny{sol}}}\,{\cal{J}}_{\mbox{\tiny{sol}}}\right)-\alpha_{m}\, \Phi_m {\cal Q}_m \right]^{\frac{z}{z+d_{\mbox{\tiny{eff}}}}}\left({\cal{M}}-\alpha_{e}\, \Phi_e {\cal Q}_e-\Omega {\cal{J}}\right)
^{\frac{d_{\mbox{\tiny{eff}}}}{z+d_{\mbox{\tiny{eff}}}}},
\end{eqnarray}
which represents the Cardy-like formula proposed in this work for charged and rotating solitons and BHs. It generalizes the standard Cardy expression by incorporating the dynamical exponent $z$ and the effective spatial
dimensionality $d_{\text{\tiny eff}}$, which control the entropy–temperature scaling given previously in Eq. (\ref{eq:S-T}). Here, the thermodynamical quantities  ${\cal{M}}_{\mbox{\tiny{sol}}}$, $\Omega_{\mbox{\tiny{sol}}}$, and ${\cal{J}}_{\mbox{\tiny{sol}}}$ encode the mass, angular velocity, and angular momentum for the soliton, obtained by a double Wick rotation (\ref{eq:doublewick}). Likewise, $\Phi_m$ and ${\cal{Q}}_m$ represent the magnetic potential and magnetic charge, and $\alpha_m$ is the counterpart of $\alpha_e$ for the magnetic sector \footnote{Here it is important to note that for Ref.  \cite{Bravo-Gaete:2015iwa}, $\alpha_e=-\alpha_m=\alpha$.}. Additionally, the exponents $z/(z+d_{\text{\tiny eff}})$ and $d_{\text{\tiny eff}}/(z+d_{\text{\tiny eff}})$ ensure homogeneity under the scaling transformations given by the first law (\ref{firstlawrotating}) and the Smarr relation (\ref{smarrrotating}). This structure guarantees that the entropy obtained from the Cardy-like formula (\ref{Cardyrotatinghvm}) matches the entropy for all examples analyzed in this paper, providing a strong consistency check of the proposed expression. In isotropic, anisotropic, or hyperscaling-violating cases and with neutral non-rotating solutions, Eq.~(\ref{Cardyrotatinghvm}) reduces to the familiar Cardy-like expressions in the literature (see, for example, Refs. \cite{Bravo-Gaete:2015wua,Shaghoulian:2015lcn,Bravo-Gaete:2015iwa,Ayon-Beato:2019kmz}).}

The novelty of the present work lies in the formulation of a Cardy-like entropy expression \eqref{Cardyrotatinghvm} that applies to charged and/or rotating BHs, as well as to their corresponding solitonic configurations, where to recover the semiclassical entropy requires the soliton to be charged and/or rotating. 

The rest of the paper is organized as follows: In Section \ref{action,eq,sol}, we will present the general derivation of the spinning configuration, which is derived from the static one through a coordinate transformation of the time and angular coordinates, as well as their thermodynamic parameters. In the next section, we explore a concrete case of a \textcolor{black}{novel} rotating charged  BH, where to recover the semiclassical entropy, the corresponding soliton must possess thermodynamic quantities beyond its mass. In Section \ref{anisotropicsection}, we consider hyperscaling violation rotating BHs \textcolor{black}{ and in Section \ref{sec:3Drot}, we show the consistency of the expression (\ref{Cardyrotatinghvm}) with a concrete example where a Lorentz boost is not required.} 
Finally, Section \ref{conclusions}  is devoted to conclusions and discussions.

\section{Deriving the general configuration from a static one}\label{action,eq,sol}

As was shown in the introduction, the present section aims to consider a general expression for the rotating configurations, which are obtained from the static ones. Here, considering the generalized metric for a {$d-$dimensional space-time}
\begin{equation}
 ds^2 = -G(r) dt^{2}+\frac{dr^{2}}{F(r)}+R^2(r) d\varphi^2{+r^{\frac{2(d-2-\theta)}{d-2}} \sum_{i=1}^{d-3} dx_i^2,}
 \label{metricstatic}
\end{equation}
together with a Lorentz boost in the $(t,\varphi)-$plane given by
\begin{equation} t\to
\frac{1}{\sqrt{1-w^2}}(t+w \varphi),\,\, \varphi\to
\frac{1}{\sqrt{1-w^2}}(\varphi+w t),\label{boost}
\end{equation}
which is well-defined for $w^2<1$, yields a rotating metric Ansatz given by
\textcolor{black}{\begin{eqnarray}
 ds^2 &=& -N(r) dt^{2}+\frac{dr^{2}}{F(r)}+H^{2}(r)\,(d\varphi+N^{\varphi}(r)dt)^{2}+{r^{\frac{2(d-2-\theta)}{d-2}} \sum_{i=1}^{d-3} dx_i^2,}
 \label{metricrotant}
\end{eqnarray}
with}
\begin{eqnarray}
N(r)&=&\frac{ R^2(r)  G(r) (1-w^2)}{R^2(r)-w^2 G(r)},\label{eq:rotatingfunctions}\\ N^{\varphi}(r)&=&\frac{w (R^2(r)-G(r))}{R^2(r)-w^2 G(r)},\nonumber\\
H^2(r)&=&\frac{R^2(r)-w^2 G(r)}{1-w^2},\nonumber
\end{eqnarray}
while the respective spinning soliton obtained through to the double Wick rotation \begin{equation}\label{eq:doublewick}
t \rightarrow i \varphi, \quad \varphi \rightarrow i t,
\end{equation} takes the form
\textcolor{black}{\begin{eqnarray}
 d{s}^2 &=& -H^{2}({r})\,\left({{d{t}}}+N^{\varphi}({r})
 \,d{\varphi}\right)^{2}+
 \frac{d{r}^{2}}{F({r})}+N({r})  d{\varphi}^{2}+{r^{\frac{2(d-2-\theta)}{d-2}} \sum_{i=1}^{d-3} dx_i^2.}
 \label{rotmetricsoliton}
\end{eqnarray}
In} all the examples that we will see, the Lagrangian ${\cal{L}}$ will be expressed in terms of the metric $g_{\mu \nu}$ and the matter fields, represented as $\Phi_{i}$, this is
\begin{equation}
{{\cal{L}}={\cal{L}}(g_{\mu \nu}, R_{\mu \nu \sigma \rho},\Phi_{i})}, \label{eq:genLagrangian}
\end{equation}
while that the corresponding action reads
\begin{equation}
S=\int d^{d}x \sqrt{-g} {\cal{L}}.\label{eq:genaction}
\end{equation}

Regarding thermodynamics, to corroborate the spinning Cardy-like expression (\ref{Cardyrotatinghvm}) with the entropy of these solutions, we will consider the Wald formula \cite{Wald:1993nt}. Using the general metric Ansatz (\ref{metricrotant})-(\ref{eq:rotatingfunctions}), this is expressed as follows:
\begin{eqnarray}
 {\cal{S}}_{W} &=& -2 \pi \oint_{\Sigma} d ^{d-2}x \sqrt{|\gamma|} P^{\alpha \beta \gamma \delta} \, \varepsilon_{\alpha \beta} \,  \varepsilon_{\gamma \delta}\nonumber\\ 
 &-& 2 \pi \Sigma_{d-2} \, {r_h^{d-3}}\,H(r_h) \,
 \Big[ P^{\alpha \beta \gamma \delta} \, \varepsilon_{\alpha \beta } \,  \varepsilon_{\gamma \delta} \Big]_{r = r_{h}}, \label{eq:Sw}
\end{eqnarray}
where $P^{\alpha \beta \gamma \delta}={\partial \mathcal{L}}/{\partial R_{\alpha \beta \gamma \delta}}$ and the expression is evaluated at the spatial section $\Sigma$ of the event horizon, while that $\Sigma_{d-2} = \int d\varphi dx_1 dx_2 \cdots dx_{d-3}=2 \pi \sigma_1 \sigma_2\cdots \sigma_{d-3}$. The term $|\gamma|$ represents the determinant of the induced metric on $\Sigma$, while $\varepsilon_{\alpha \beta}$ denotes the binormal vector
$$\varepsilon_{\alpha \beta}=-\varepsilon_{ \beta \alpha}:=\frac{1}{\kappa}\,\nabla_{\alpha} \xi_{\beta},$$
where  $\kappa$ is the surface gravity
\begin{equation*}\label{kappa}
\kappa=\sqrt{-\frac{1}{2}\left(\nabla_{\mu} \xi_{\nu}\right)\left(\nabla^{\mu} \xi^{\nu}\right)},
\end{equation*}
with a Killing vector $\xi^{\mu}=(\partial_t)^{\mu}-\Omega (\partial_\varphi)^{\mu}$, and the angular velocity $\Omega$  given by  
\begin{equation}\label{omega}
\Omega=N^{\varphi}(r) \Big{|}_{r=r_h}={w}.
\end{equation}
The Hawking temperature $T$ reads
\begin{equation}\label{eq:Thawking}
T=\frac{\kappa}{2\pi}=\frac{G'(r)}{4 \pi}\,\sqrt{\frac{F(r) (1-w^2)}{G(r)}}\,\,\Bigg{|}_{r=r_h},
\end{equation}
where $(')$ denotes the derivative with respect to the radial coordinate $r$, and $r_h$ the location of the event horizon. On the other hand, we will compute its thermodynamic parameters as well as its respective BH solution by using the formulation proposed by Kim et al. \cite{Kim:2013zha,Gim:2014nba}, where the idea corresponds to expressing the ADT potential  {on-shell} $Q_{\mbox{\tiny{ADT}}}^{\mu\nu}$  \cite{Deser:2002rt,Deser:2002jk}
$$
\sqrt{-g}\,Q_{\mbox{\tiny{ADT}}}^{\mu\nu}=\frac{1}{2}\delta
K^{\mu\nu}-\xi^{[\mu}\Theta^{\nu]},
$$
through a Killing vector $\xi^{\mu}\partial_{\mu}$, a Noether potential off-shell $K^{\mu\nu}$ together with  and a surface term $\Theta^{\mu}$ obtained from the variation of the action (\ref{eq:genaction}). With all the above, and by using a uni parametric path inside the solution space, characterized by a constant  $s \in [0,1]$ and interpolating between the solution and a base solution, we can find the conserved charge $Q(\xi)$:
\begin{equation}
\label{chargequasi} Q(\xi)\!=\!\int_{\cal B}\!
{dx^{d-2}_{\mu\nu}}\Big(\delta K^{\mu\nu}(\xi)-2\xi^{[\mu} \!\! \int^1_0ds~
\Theta^{\nu]}\Big),
\end{equation}
where $\delta K^{\mu\nu}(\xi) =
K^{\mu\nu}_{s=1}(\xi)-K^{\mu\nu}_{s=0}(\xi)$ is the difference of the Noether potential between both solutions, while {$dx^{d-2}_{\mu\nu}$} represents the integration over the two-dimensional boundary ${\cal B}$.

It is important to note that the quasilocal formalism has been a powerful tool to compute the mass ${\cal{M}}$ in a correct way, with a Killing vector of the form {$\partial_t = \xi^{\mu} \partial_{\mu}$}, of BHs with non-standard asymptotic behavior  \cite{Ayon-Beato:2015jga, Ayon-Beato:2019kmz,Bravo-Gaete:2021kgt,Bravo-Gaete:2025lgs}, as well as the mass ${\cal{M}}_{\tiny{\mbox{sol}}}$ of its respective solitonic configurations. In addition, it has been shown that this formalism allows to compute the angular momentum ${\cal{J}}$ of spinning BH solutions with an election of the Killing vector given by {$\partial_\varphi = \xi^{\mu} \partial_{\mu}$}   \cite{Kim:2013cor,Kim:2013qra,Peng:2015yjx}. Furthermore, as we will discuss below, it also calculates the angular momentum (${\cal{J}}_{\tiny{\mbox{sol}}}$) of spinning solitonic configurations.

With all these components, we are now in a position to show the validity of the Cardy-like expression (\ref{Cardyrotatinghvm}) through a variety of concrete examples. The first {case will be the isotropic case, given by a charged spinning configuration where, to recover the semiclassical entropy, the magnetic charge and the angular momentum of the soliton are principal components.}

\section{A new charged rotating black hole in four-dimensional Critical Gravity}\label{isotropic}

In the following lines, and as shown in the introduction, we will check the validity of the formula (\ref{Cardyrotatinghvm}) {considering the Einstein-Hilbert (EH) action together with a cosmological constant supplemented with the so-called four-dimensional Critical Gravity (CG) \cite{Lu:2011zk}:
\textcolor{black}{\begin{eqnarray}\label{eq:action-EHLambda-CG}
\mathcal{L}_{\tiny{CG}}&=&\frac{1}{2} \left(R-2\Lambda\right)+\frac{1}{2}\left[
-\left(\frac{1}{2\Lambda}\right)\,{R}^2+\left(\frac{3}{2\Lambda}\right){R}_{\mu \nu}{R}^{\mu \nu}\right].
\end{eqnarray}
The} resulting theory is both ghost-free and renormalizable, incorporating quadratic curvature corrections, albeit at the cost of fourth-order equations of motion and yielding a BH configuration with vanishing extensive thermodynamic quantities. Nevertheless, in Refs. \cite{Alvarez:2022upr,Lin:2024ubg}, this limitation can be circumvented by the addition of an appropriate matter source via nonlinear electrodynamics in the Plebánski formalism~\cite{Plebanski:1968}, thereby yielding the first instance in four-dimensional CG in which the BH thermodynamic parameters are non-vanishing.
Concretely, the nonlinear electrodynamics structure, given by the matter Lagrangian $\mathcal{L}_{\tiny{\mbox{matter}}}$, takes the form
\begin{eqnarray}\label{eq:NLE}
\mathcal{L}_{\tiny{\mbox{matter}}} = -\frac{1}{2} F_{\mu\nu} P^{\mu\nu} + \mathcal{H}(P),
\end{eqnarray}
where $F_{\mu\nu} := 2\partial_{[\mu} A_{\nu]}$ is the usual field strength tensor, and $P^{\mu\nu}$ denotes its conjugate antisymmetric tensor. The scalar invariant $P$ is defined as $P = \frac{1}{4} P_{\mu\nu} P^{\mu\nu}$, and $\mathcal{H}(P)$ encodes the nonlinear dynamics of the electromagnetic field. }

{Considering the Lagrangian (\ref{eq:genLagrangian}) given by ${\cal L}=\mathcal{L}_{\tiny{CG}}+\mathcal{L}_{\tiny{\mbox{matter}}}$, and the structural function $\mathcal{H}(P)$ as
\textcolor{black}{\begin{eqnarray}\label{eq:H(P)}
\mathcal{H}(P) &=& \frac{1}{3}(\alpha_2^2 - 3\alpha_1\alpha_3) P - 2\alpha_1 (-2P)^{1/4} + \alpha_2 \sqrt{-2P},
\end{eqnarray}
the} solution, with $d=4$ and $\theta=0$, is given by eqs. 
(\ref{metricrotant})-(\ref{eq:rotatingfunctions}), where
\begin{eqnarray*}\label{frotatingCG}
G(r)&=&F(r)=r^2 \left(1 - \frac{\alpha_1 \sqrt{M}}{r} + \frac{\alpha_2 M}{r^2} - \frac{\alpha_3 M^{3/2}}{r^3} \right),\nonumber\\
R^2(r)&=&r^2,
\end{eqnarray*}
while the potential takes the form $A_{\mu} dx^{\mu}=\frac{1}{\sqrt{1-w^2}}A_{t}(r) \left(dt+w d \varphi\right)$, where
\begin{eqnarray*}\label{frotatingCG}
A_t(r)=\frac{\alpha_1 r^2}{2\sqrt{M}} - \alpha_2 r + \frac{M(3\alpha_1 \alpha_3 - \alpha_2^2)}{3r},
\end{eqnarray*}
and the cosmological constant is 
\begin{equation}\label{eq:Lambda1}
\Lambda=-3.
\end{equation}}

{The thermodynamic quantities associated with this configuration can be computed explicitly. The electric charge and electric potential read (following Refs. \cite{Herrera-Aguilar:2021top,Dehghani:2015gza,Sheykhi:2016tma}): 
\textcolor{black}{\begin{eqnarray}
{\cal{Q}}_e &=& \frac{r_h^2 \Sigma_2}{\zeta^2 \sqrt{1-w^2}}, \quad
\Phi_e = r_h \sqrt{1-w^2} \Big( \alpha_2 + \alpha_1^2 - \frac{3}{2} \alpha_1 \zeta - \frac{\alpha_1 \alpha_2}{\zeta} + \frac{1}{3} \frac{\alpha_2^2}{\zeta^2} \Big),
\end{eqnarray}
while} the expressions for the mass, Hawking temperature, angular momentum, and entropy are given by
\begin{eqnarray}
\cal{M} &=& \frac{\alpha_1 \alpha_2 r_h^3 (2+w^2)  \Sigma_2}{18 \zeta^3 (1-w^2)}, \\
T &=& \frac{r_h \sqrt{1-w^2}}{4\pi} \left(3 - \frac{2\alpha_1}{\zeta} + \frac{\alpha_2}{\zeta^2} \right), \label{eq:T-1} \\
\cal{J}&=&\frac{ \alpha_1 \alpha_2 w r_h^3 \Sigma_2}{6 \zeta^3 (1-w^2)}, \\
{\cal{S}}_{W} &=&\frac{ 2\pi \Sigma_2 r_h^2}{\sqrt{1-w^2}} \left( \frac{\alpha_1}{\zeta} - \frac{2\alpha_2}{3\zeta^2} \right). \label{eq:ent1}
\end{eqnarray}
Here,  $r_h$ denotes the horizon radius, which is parametrized as $r_h = \zeta \sqrt{M}$, where $\zeta$ satisfies the cubic equation
\begin{equation}
\zeta^3 - \alpha_1 \zeta^2 + \alpha_2 \zeta - \alpha_3 = 0,
\label{eq:cubic}
\end{equation}
and} these expressions satisfy the first law of BH thermodynamics
\begin{equation}\label{firstlawrotating}
\delta{\cal{M}}=T \delta{\cal{S}}+\Omega \delta{\cal{J}}+\Phi_e \delta {\cal Q}_e.
\end{equation}
\textcolor{black}{It is straightforward to check that upon the following dimensional scaling $r_h \rightarrow \tilde{\lambda} r_h$, we have:
$${\cal{M}} \rightarrow \tilde{\lambda}^3 {\cal{M}},\quad {\cal{S}}_{W} \rightarrow \tilde{\lambda}^2 {\cal{S}}_{W},\quad {\cal{J}} \rightarrow \tilde{\lambda}^3 {\cal{J}},\quad {\cal{Q}}_{e} \rightarrow \tilde{\lambda}^2 {\cal{Q}}_{e},$$
and the Smarr formula (\ref{smarrrotating}) is recovered} with $d_{\tiny{\mbox{eff}}}=2, z=1$, and  $\alpha_e=2/3$, 
with the angular velocity $\Omega$  given in (\ref{omega}).

{It is important to note that, as a first attempt to reproduce the entropy (\ref{eq:ent1}) from a Cardy-like expression, is considering, as shown \cite{BravoGaete:2017dso,Shaghoulian:2015lcn}, a solitonic uncharged non-spinning configuration (via the transformation (\ref{eq:doublewick}) considering $\alpha_1=\alpha_2=w=0$), given by the line element (\ref{rotmetricsoliton}):
\textcolor{black}{\begin{eqnarray}
F(r)&=&N(r)=r^2\left(1-\left(\frac{r_0}{r}\right)^3\right), \quad H^2(r)=r^2,\qquad N^{\varphi}(r)=0,\nonumber\\
\end{eqnarray}
with} $r_0\leq r < +\infty$. Additionally, the regularity of the spacetime requires that the coordinates $\varphi$ be periodic, with period 
$$\frac{3 r_0}{4 \pi}=\frac{1}{2 \pi} \Rightarrow r_0=\frac{2}{3}.$$
Nevertheless, the quasilocal formulation \cite{Kim:2013zha,Gim:2014nba} reveals that the mass of the soliton vanishes and the entropy (\ref{eq:ent1}) cannot be recovered. From the above, a natural extension is to consider a solitonic configuration with a richer thermodynamic structure. One of them is via a spinning magnetically charged solution, which is given by eqs. (\ref{eq:rotatingfunctions}) and (\ref{rotmetricsoliton}):
\textcolor{black}{\begin{eqnarray}\label{frotatingCGsol}
G(r)&=&F(r)=r^2 \left(1 - \frac{\alpha_1 \sqrt{\mu}}{r} + \frac{\alpha_2 \mu}{r^2} - \frac{\alpha_3 \mu^{3/2}}{r^3} \right), \mbox{ with } \mu=\left(\frac{r_0}{\zeta}\right)^2,\nonumber\\
R^2(r)&=&r^2,
\end{eqnarray}
and a} potential $A_{\mu} dx^{\mu}=\frac{1}{\sqrt{1-w^2}}A_{\varphi}(r) \left(w dt+d \varphi\right)$, where
\begin{eqnarray}\label{eq:Aphi}
A_\varphi(r)=\frac{\alpha_1 r^2}{2\sqrt{\mu}}- \alpha_2 r+\frac{\mu(3\alpha_1 \alpha_3 - \alpha_2^2)}{3r}. 
\end{eqnarray}
The structural function $\mathcal{H}$ is obtained from eq. (\ref{eq:H(P)}) through the change in the invariant $P$ for $−P$ and the cosmological constant is given by (\ref{eq:Lambda1}). As before, $\zeta$ is a constant satisfying eq. (\ref{eq:cubic}) and, to ensure the correct identification with its Euclidean version
\begin{equation}\label{eq:r0}
r_0=\frac{2}{\left(3-\frac{2\alpha_1}{\zeta} + \frac{\alpha_2}{\zeta^2}\right)\sqrt{1-w^2}}.
\end{equation}
For this newly obtained magnetically charged rotating configuration, the absence of an integration constant, and thus of an event horizon, implies that the solution described by eqs. (\ref{eq:rotatingfunctions}), (\ref{rotmetricsoliton}), and (\ref{frotatingCGsol})-(\ref{eq:r0}) does not possess entropy. Nonetheless, it remains possible to compute its mass ${\cal{M}}_{\mbox{\tiny{sol}}}$ and angular momentum ${\cal{J}}_{\mbox{\tiny{sol}}}$ via the formalism \cite{Kim:2013zha,Gim:2014nba}, which read
\begin{eqnarray}\label{eq:massand-sol1}
{\cal{M}}_{\mbox{\tiny{sol}}}&=& -\frac{\alpha_1 \alpha_2  \Sigma_2 (1+2w^2)}{18 (1-w^2)} \left(\frac{r_0}{\zeta}\right)^{3},\\
{\cal{J}}_{\mbox{\tiny{sol}}}&=&\frac{\alpha_1 \alpha_2  w \Sigma_2}{6 (1-w^2)} \left(\frac{r_0}{\zeta}\right)^{3}.
\end{eqnarray}
In addition, we can note that this charged rotating soliton has magnetic charge ${\cal{Q}}_{m}$ and potential $\Phi_{m}$, which take the form
\textcolor{black}{\begin{eqnarray}
{\cal{Q}}_m &=& \frac{\Sigma_2 r_0^2}{\zeta^2 \sqrt{1-w^2}}, \quad
\Phi_m = r_0 \sqrt{1-w^2} \Big( \alpha_2 + \alpha_1^2 - \frac{3}{2} \alpha_1 \zeta - \frac{\alpha_1 \alpha_2}{\zeta} + \frac{1}{3} \frac{\alpha_2^2}{\zeta^2} \Big),\label{eq:phim}
\end{eqnarray}
with} $r_0$ given previously in (\ref{eq:r0}), and the angular velocity reads
\begin{eqnarray}\label{eq:angu-vel-sol}
\Omega_{\mbox{\tiny{sol}}}=w.
\end{eqnarray}
Finally, with all these quantities, the entropy (\ref{eq:ent1}) can be recovered from the Cardy-like expression (\ref{Cardyrotatinghvm}) with  $d_{\tiny{\mbox{eff}}}=2, z=1$, $\alpha_e=\alpha_m=2/3$. \textcolor{black}{This is, ${\cal{S}}_{C}={\cal{S}}_{W}.$}}

Building on the above, in the next section we proceed to present additional examples of configurations exhibiting non-standard asymptotic behavior, focusing on the spinning case (this is, ${
\cal Q}_{m}=0={\cal Q}_e$). 

\section{Studying spinning Hyperscaling violation configurations}\label{anisotropicsection}

To provide more concrete examples {to show the consistency of the expression (\ref{Cardyrotatinghvm}) with configurations with other asymptotic behaviors}, we will consider {uncharged spinning} hyperscaling violation BHs described by the metric in eq. (\ref{eq:Hyperscaling}), characterized by the presence of the hyperscaling violation exponent $\theta$. As a first case,  we consider the { Lagrangian: \begin{eqnarray}
\mathcal{L}&=&\frac{R}{2} -\left(\frac{1}{2}\,\nabla^{\mu} \phi \nabla_{\mu}
\phi+U(\phi)\right),\label{ActionCFM}
\end{eqnarray}
this} is the Einstein gravity together with a
scalar field minimally coupled and a potential $U(\phi)$ given by \cite{BravoGaete:2017dso,Perlmutter:2012he}:
\begin{eqnarray}
\phi(r)&=&\sqrt {\frac{\theta\, \left( \theta-d+2 \right)
}{d-2}}\ln(r), \nonumber\\
U(\phi)&=&-\frac{\left( d-2-\theta \right)  \left(
d-1-\theta\right)}{2} \,{e}^{\frac{2
\sqrt{\theta}\,\phi}{\sqrt{(\theta-d+2)(d-2)}}}, \label{scapotentialES}
\end{eqnarray}
where the spinning configuration obtained through the coordinate transformation (\ref{boost}) with $z=1$, is given by the line element (\ref{metricrotant})-(\ref{eq:rotatingfunctions}), where
\begin{eqnarray*}\label{hvmrotatingES}
G(r)&=& {r}^{\frac{2(d-2-\theta)}{d-2}}\, \left[1-\left(\frac{r_h}{r}\right)^{d-1-\theta}\right]
, \quad
R^2(r)=r^{\frac{2(d-2-\theta)}{d-2}},\nonumber\\ 
F(r)&=& {r}^{\frac{2(d-2+\theta)}{d-2}} \left[1-\left(\frac{r_h}{r}\right)^{d-1-\theta}\right].
\end{eqnarray*}
As before, $r_h$ corresponds to the location of the event horizon, and the expressions for the respective
thermodynamic parameters are:
\begin{eqnarray}
{\cal{S}}_{W}&=& \frac{2 \,\pi\, \Sigma_{d-2} r_{h}^{d-2-\theta}}{\sqrt{1-w^{2}}},\label{waldcubicrotES}\\ 
T&=& \frac{r_h(d-1-\theta)\sqrt{1-w^{2}}}{4 \pi},\\
{\cal{M}}&=&{\frac { \left( w^{2}+d-2-\theta\right)r_{h}^{d-1-\theta} \Sigma_{d-2}}{ 2 \left(
1-{w}^{2} \right) }},
\label{masscubicrotES}  \\
{\cal{J}}&=&{\frac {\left( d-1-\theta \right) w r_{h}^{d-1-\theta} \Sigma_{d-2}}{2 \left(
1-{w}^{2} \right)}}.\label{momcubicrotES}
\end{eqnarray}
Here, we verify that the first law of thermodynamics (\ref{firstlawrotating}) \textcolor{black}{holds. As in the previous case, we note that scaling the location of the event horizon as $r_h \rightarrow \tilde{\lambda} r_h$ yields
$${\cal{M}} \rightarrow \tilde{\lambda}^{d-1-\theta} {\cal{M}},\quad {\cal{S}}_{W} \rightarrow \tilde{\lambda}^{d-2-\theta} {\cal{S}}_{W},\quad {\cal{J}} \rightarrow \tilde{\lambda}^{d-1-\theta} {\cal{J}},$$
and the Smarr relation (\ref{smarrrotating}) for 
\begin{eqnarray}\label{eq:HVMc1}
d_{\mbox{\tiny{eff}}}=d-2-\theta,
\end{eqnarray}
is satisfied with} this solution.

The corresponding rotating soliton is given by {eqs. (\ref{eq:rotatingfunctions}) and (\ref{rotmetricsoliton}):}
\begin{eqnarray*}\label{hvmrotatingESsoliton}
G(r)&=& {r}^{\frac{2(d-2-\theta)}{d-2}}\, \left[1-\left(\frac{2}{ (d-1-\theta) \sqrt{1-w^{2}}\,{r}}\right)^{d-1-\theta}\right]
, \\
F(r)&=& {r}^{\frac{2(d-2+\theta)}{d-2}} \left[1-\left(\frac{2}{ (d-1-\theta) \sqrt{1-w^{2}}\,{r}}\right)^{d-1-\theta}\right],\\
R^2(r)&=&r^{\frac{2(d-2-\theta)}{d-2}},
\end{eqnarray*}
where the scalar field and potential are given in  (\ref{scapotentialES}). As previously, by using the quasilocal formalism, the mass and angular momentum
of the spinning soliton read
\textcolor{black}{\begin{eqnarray}
{\cal{M}}_{\mbox{\tiny{sol}}}&=&-\frac { \left( 1+w^{2}(d-2-\theta)
\right) \Sigma_{d-2}}{2
 \left( 1-{w}^{2} \right) }\left[ \frac
{2}{\sqrt {1-{w}^{2}} \left( d-1-\theta\right)}
 \right]^{d-1-\theta}\label{masssolcubicrotES}
,\\ {\cal{J}}_{\mbox{\tiny{sol}}}&=& {\frac { \left( d-1-\theta \right) w \Sigma_{d-2}}{ 2\left( 1-w^{2} \right) }}\left[ {\frac {2}{\sqrt
{1-{w}^{2}} \left( d-1-\theta\right) }}
 \right] ^{d-1-\theta}.\label{momsolcubicrotES}
\end{eqnarray}
Finally,} it is straightforward to check that the formula
(\ref{Cardyrotatinghvm}) with $d_{\mbox{\tiny{eff}}}$ {given in (\ref{eq:HVMc1}) and the angular momentum (\ref{omega})} fits perfectly with the expressions of the mass and
angular momentum (\ref{masscubicrotES})-(\ref{momcubicrotES}), its
soliton counterpart
(\ref{masssolcubicrotES})-(\ref{momsolcubicrotES}) with $\Omega_{\mbox{\tiny{sol}}}=w$, and the entropy
(\ref{waldcubicrotES}).


As discussed in the introduction, the expression of the effective spatial dimensionality $d_{\mbox{\tiny{eff}}}$ is not universal and depends on the considered theory. It is for this reason that, to corroborate the Cardy-like formula (\ref{Cardyrotatinghvm}) in more concrete solutions, we consider as a pure quadratic gravity theory given by \cite{BravoGaete:2017dso}
\begin{eqnarray}
\mathcal{L}&=&\frac{1}{2}
\left(\beta_1{R}^2 +\beta_2{R}_{\alpha\beta}{R}^{\alpha\beta}
\right). \label{eq:Squadhvm}
\end{eqnarray}
The spinning configuration (with the metric Ansatz (\ref{metricrotant}) with (\ref{eq:rotatingfunctions})), $\theta=d-1$ and $z=1$, is given by
\begin{eqnarray*}\label{hvmrotatingsol1}
G(r)&=& r^{-\frac{2}{d-2}}\, \left[1-\left(\frac{r_h}{r}\right)^{\frac{2(d-1)}{d-2}}\right]
, \quad
R^2(r)=r^{-\frac{2}{d-2}},\nonumber\\ 
F(r)&=& {r}^{\frac{2(2d-3)}{d-2}} \left[1-\left(\frac{r_h}{r}\right)^{\frac{2(d-1)}{d-2}}\right],
\end{eqnarray*}
together with $\beta_1=-\frac{(d+2)\beta_2}{(5d-2)}$. In this case, the thermodynamic quantities read
\begin{eqnarray}
{\cal{S}}_{W}&=&{\frac {16 \pi {\beta_2} (d-1)^2 \Sigma_{d-2} r_h^{\frac{d}{d-2}}}{ (5d-2)(d-2)\sqrt {1-w^{2}}}}
 ,\\
 T&=&{\frac {(d-1) \sqrt {1-w^{2}}\,r_h}{2 (d-2)  \pi }},\label{enttemphvmquad}\\
{\cal{M}}&=&{\frac {4 {\beta_2} \left( d+(d-2) w^{2}\right) (d-1)^2 \Sigma_{d-2}  
r_h^{\frac{2(d-1)}{d-2}}}{  (d-2)^2 (5d-2)\left(1- w^{2} \right) }}
 ,\label{masshvmquad}\\
 {\cal{J}}&=&{\frac {8 {\beta_2} (d-1)^3 w \Sigma_{d-2} 
r_h^{\frac{2(d-1)}{d-2}}}{  (d-2)^2 (5d-2)\left(1- w^{2} \right) }},\label{anghvmquad}
\end{eqnarray}
where with $\Omega$ as eq. (\ref{omega}), the first law (\ref{firstlawrotating}) \textcolor{black}{is satisfied. As before, via the scaling of $r_h \rightarrow \tilde{\lambda} r_h$, we obtain  
$${\cal{M}} \rightarrow \tilde{\lambda}^{2(d-1)/(d-2)} {\cal{M}},\quad {\cal{S}}_{W} \rightarrow \tilde{\lambda}^{d/(d-2)} {\cal{S}}_{W},\quad {\cal{J}} \rightarrow \tilde{\lambda}^{2(d-1)/(d-2)} {\cal{J}}.$$
Recovering the Smarr formula (\ref{smarrrotating}) with
\begin{equation}\label{eq:deff-cuad}
d_{\mbox{\tiny{eff}}}=\frac{d}{d-2}.
\end{equation}
The} thermodynamic components of its solitonic spinning counterpart take the form
\textcolor{black}{\begin{eqnarray*}
{\cal{M}}_{\mbox{\tiny{sol}}}&=&-\frac{16 \beta_2 \big((d-2)+d w^2\big) \Sigma_{d-2}}{(d-2)(5d-2)(1-w^2)^{\frac{2d-3}{d-2}}}\left(\frac{d-2}{4}\right)^{\frac{d}{d-2}}\left(\frac{4}{d-1}\right)^{\frac{2}{d-2}}
,\\
{\cal{J}}_{\mbox{\tiny{sol}}}&=& \frac{32 \beta_2 (d-1) w \Sigma_{d-2}}{(5d-2)(d-2) (1-w^2)^{\frac{2d-3}{d-2}}}\left(\frac{d-2}{4}\right)^{\frac{d}{d-2}}\left(\frac{4}{d-1}\right)^{\frac{2}{d-2}},
\end{eqnarray*}
whose} line element, written in (\ref{eq:rotatingfunctions})-(\ref{rotmetricsoliton}), has the following expression for the metric functions
\begin{eqnarray*}\label{hvmrotatingsol1soliton}
G(r)&=& r^{-\frac{2}{d-2}}\, \left[1-\left(\frac{d-2}{(d-1)\sqrt{1-w^2}\,r}\right)^{\frac{2(d-1)}{d-2}}\right]
,\\
R^2(r)&=&r^{-\frac{2}{d-2}},\nonumber\\ 
F(r)&=& {r}^{\frac{2(2d-3)}{d-2}} \left[1-\left(\frac{d-2}{(d-1)\sqrt{1-w^2}\,r}\right)^{\frac{2(d-1)}{d-2}}\right].
\end{eqnarray*}
As in the previous situations, it is straightforward to check that the Cardy-like expression (\ref{Cardyrotatinghvm}), with $d_{\mbox{\tiny{eff}}}$ given in (\ref{eq:deff-cuad}), correctly fixes the Wald entropy obtained in (\ref{enttemphvmquad}), where $\Omega_{\mbox{\tiny{sol}}}$ takes the form given previously in eq. (\ref{eq:angu-vel-sol}).

\textcolor{black}{
\section{Three-dimensional rotating Black Hole with scalar hair}
\label{sec:3Drot}}

\textcolor{black}{To strengthen the robustness of the expression (\ref{Cardyrotatinghvm}), it is instructive to analyze spinning configurations that do not originate from the coordinate transformation (\ref{boost}). In particular, we can consider a rotating BH solution endowed with scalar hair. This configuration was obtained in \cite{Xu:2014uha,Zou:2014gla} within three-dimensional Einstein gravity nonminimally coupled to a scalar field $\phi$. The corresponding action reads:
\begin{eqnarray}
&&\mathcal{L}=\frac{1}{2}\big(
R-2\Lambda \big)-\Biggl[
\frac{1}{2}\nabla_{\mu}\phi\nabla^{\mu}\phi+\frac{1}{16}R\phi^2+U(\phi)
\Biggr], \label{actionscalarfield}
\end{eqnarray}
where the cosmological constant is set to $\Lambda = -1$, and the scalar potential \( U(\phi) \) is given by:
\begin{eqnarray}
U(\phi) = \frac{{\phi}^{10} \left({\phi}^{6} - 40\,{\phi}^{4} - 4608 + 640\,{\phi}^{2} \right)\left(\alpha-1\right)}{512 \left(3\eta+2\right)^{2}\left( \kappa{\phi}^{2}-8 \right) ^{5}} + \frac{1}{512} \left( 1 - \frac{\alpha \eta^3}{3\eta+2} \right) {\phi}^{6}.
\end{eqnarray}
This potential is parameterized by the constants \(\eta\) and \(\alpha\), and the scalar field \(\phi\) is expressed as:
\begin{eqnarray}
\phi(r) = \pm \sqrt{\frac{8 r_{h}}{\eta r + r_{h}}},
\end{eqnarray}
and the corresponding rotating BH solution, described by the line element (\ref{metricrotant}) with $d=3$ and $\theta=0$, takes the form:
\begin{eqnarray}
N(r) = F(r) = r^{2} - \frac{\alpha\,r_h^2\,(3\,\eta r+2 r_h)}{(3\eta+2) r} + \frac{r_h^4 (3\eta r+2 r_h)^2 \big(\alpha-1\big)}{(3\eta+2)^2 r^4},\label{eq:F}
\end{eqnarray}
\begin{eqnarray}
H^2(r) = r^{2}, \quad N^{\varphi}(r) = -\frac{\sqrt{\alpha-1} (3\eta r+2 r_h) r_h^2 }{(3\eta+2) r^3}, \label{rotscalarfieldsoln}
\end{eqnarray}
where $r_h$ denotes the event horizon radius. This solution is interesting because it incorporates a non-trivial scalar field configuration, which modifies the spacetime geometry and thermodynamic properties of the BH configuration, where their parameters are computed as follows
\begin{eqnarray}
{\cal{S}}_{W} = \frac{4 \eta \pi^{2} r_h}{1+\eta}, \quad T = \frac{3 (\eta+1) (2-\alpha) r_h}{2 \pi (3\eta+2)},\label{entroptempsol1}
\end{eqnarray}
\begin{eqnarray}
{\cal M} = \frac{3 \alpha r_h^{2} \pi \eta}{3\eta+2}, \quad {\cal J} = -\frac{6 \pi \sqrt{\alpha-1} \eta r_h^2}{3\eta+2},\label{entroptempsol2}
\end{eqnarray}
together with an angular momentum $\Omega=N^{\varphi}(r_h)=-\sqrt{\alpha-1}
$ and satisfying the uncharged spinning first law (\ref{firstlawrotating}). Through the scaling  $r_h \rightarrow \tilde{\lambda} r_h$, we have that the extensive thermodynamical parameters are scaled as:
$${\cal{M}} \rightarrow \tilde{\lambda}^{2} {\cal{M}},\quad {\cal{S}}_{W} \rightarrow \tilde{\lambda} {\cal{S}}_{W},\quad {\cal{J}} \rightarrow \tilde{\lambda}^{2} {\cal{J}},$$  and the Smarr relation (\ref{smarrrotating}) (for $d_{\mbox{\tiny{eff}}}=z=1$) holds. For the sake of completeness, to have a real and positive expression for the thermodynamical parameters (\ref{entroptempsol1})-(\ref{entroptempsol2}), the range for the constant $\alpha$ must be $1\leq \alpha <2$ for $\eta>0$.}

\textcolor{black}{Together with the nonminimally dressed stationary BH solution (\ref{eq:F})-(\ref{rotscalarfieldsoln}), another solution with the same asymptotical behavior is given by the line element (\ref{rotmetricsoliton}) (with $d=3$ and $\theta=0$), where:
\begin{eqnarray}
F(r) = N(r) = r^{2} - \frac{\alpha\,{r}_0^2\,(3\,\eta r+2 {r}_0)}{(3\eta+2) r} + \frac{{r}_0^4 (3\eta r+2 {r}_0)^2 \big(\alpha-1\big)}{(3\eta+2)^2 r^4}, \label{soldoublewick}
\end{eqnarray}
\begin{eqnarray}
N^{\varphi}(r) = -\frac{\sqrt{\alpha-1} (3\eta r+2 {r}_0) {r}_0^2 }{(3\eta+2) r^3}, \quad H^2(r) = r^2,
\end{eqnarray}
and the scalar field is given by:
\begin{eqnarray}
\phi(r) = \pm \sqrt{\frac{8 {r}_{0}}{\eta r + {r}_{0}}},
\end{eqnarray}
while that ${r}_{0}>0$ is defined as:
\begin{eqnarray}
{r}_{0} = \frac{3\eta+2}{3(\eta+1)(2-\alpha)}.
\end{eqnarray}
Here, we note that this hairy solitonic configuration is not a BH due to the metric function $N$ present in $d \varphi ^2$,  which indicates a spinning solitonic structure. The above implies that some conditions must be considered; these are the adjustments of $r_0$ to ensure the correct identification with its Euclidean version and $r \geq r_0$. The thermodynamic quantities for the soliton configuration are
\begin{eqnarray}
{\cal{M}}_{\mbox{\tiny{sol}}} = -\frac{(3\eta+2) \alpha \eta \pi}{3\big(2-\alpha\big)^2 (\eta+1)^2},
\end{eqnarray}
\begin{eqnarray}
{\cal{J}}_{\mbox{\tiny{sol}}} = -\frac{2(3\eta+2) \eta \pi \sqrt{\alpha-1}}{3\big(2-\alpha\big)^2 (\eta+1)^2}, \quad \Omega_{\mbox{\tiny{sol}}} = -\sqrt{\alpha-1}.
\end{eqnarray}}

\textcolor{black}{Hence, both the BH and soliton sectors confirm that the expression (\ref{Cardyrotatinghvm}), with  $d_{\tiny{\mbox{eff}}}=z=1$, remains valid. This provides further evidence that the Cardy-like relation  captures the structure of the Wald entropy ${\cal S}_{W}$ given in (\ref{entroptempsol1}) for a three-dimensional rotating hairy BH beyond the class of geometries generated by the Lorentz boost (\ref{boost}). Additionally, it provides important insights into the holographic interpretation of these solutions and their significance within the broader context of the AdS/CFT correspondence \cite{Plefka:2005bk}. These concrete examples enhance our understanding of the universality of the Cardy-like formula (\ref{Cardyrotatinghvm}) and its relevance to various gravitational systems. The inclusion of these spinning configurations, both for the BH and the soliton, provides a more comprehensive understanding of the thermodynamic behavior of rotating solutions in three-dimensional gravity. The scalar field plays a crucial role in modifying the spacetime geometry and thermodynamic properties, introducing new parameters ($\eta$ and $\alpha$) that govern the dynamics of the system.  This analysis highlights the interplay between scalar fields and rotation in lower-dimensional gravity, which could have implications for understanding holographic dualities and the microstructure of these solutions.}\\

\section{Conclusions and discussions }\label{conclusions}

One of the aims of this paper is to propose a charged spinning Cardy-like formula (\ref{Cardyrotatinghvm}). This expression is characterized by the mass, charge, and angular components of the BH, along with its respective solitonic solution, which is related to the first law of thermodynamics (\ref{firstlawrotating}) as well as to the charged spinning Smarr expression (\ref{smarrrotating}).  

Alongside the above, we verify the feasibility of this formula through various concrete examples involving static BHs, whose spinning counterparts emerge via a Lorentz boost (\ref{boost}). For the sake of completeness, it is interesting to note that from a non-rotating uncharged configuration (\ref{metricstatic}) with entropy and temperature given by $\tilde{\cal{S}}$ and $\tilde{T}$ respectively, under the coordinate transformation (\ref{boost}):
\begin{eqnarray}\label{entropychange}
{\cal{S}}=\frac{\tilde{\cal{S}}}{\sqrt{1-w^2}},\qquad
T =\tilde{T}\,{\sqrt{1-w^2}},
\end{eqnarray}
where ${\cal{S}}$ and $T$ represent now the entropy and temperature of the boosted (rotating), uncharged BH.  Applying the Smarr relation (\ref{smarrrotating}), as well as the first law (\ref{firstlawrotating}) (both with ${\cal Q}_e=0$), the angular momentum ${\cal{J}}$ arises and depends on the uncharged, non-rotating mass $\tilde{\cal{M}}$ of the form
\begin{eqnarray}
{\cal{J}}=\frac{(z+d_{\tiny{\mbox{eff}}}) w \,\tilde{\cal{M}}}{d_{\tiny{\mbox{eff}}} (1-w^2)},
\end{eqnarray}
while the rotating mass becomes:
\begin{eqnarray}\label{masschange}
{\cal{M}} =\left[1+\frac{(z+d_{\tiny{\mbox{eff}}}) w^2}{d_{\tiny{\mbox{eff}}} (1-w^2)}\right] \, \tilde{\cal{M}}.
\end{eqnarray}

A similar analysis applies to the charged case. In addition to eq. (\ref{entropychange}), after the Lorentz boost (\ref{boost}), the electric charge $\tilde{\cal{Q}}_{e}$ and electric potential $\tilde{\Phi}_e$ transform as
\begin{eqnarray}\label{entropychange-charge}
{\cal{Q}}_{e}=\frac{\tilde{\cal{Q}}_e}{\sqrt{1-w^2}},\qquad
\Phi_e = \tilde{\Phi}_e\,{\sqrt{1-w^2}}.
\end{eqnarray}
The angular momentum is then given by:
\begin{eqnarray}\label{eq:jcharge}
{\cal{J}}=\frac{(z+d_{\tiny{\mbox{eff}}}) w}{d_{\tiny{\mbox{eff}}} (1-w^2)} \left[ \tilde{\cal M}+\left(\frac{d_{\tiny{\mbox{eff}}}}{z+d_{\tiny{\mbox{eff}}}}-\alpha_e\right) \tilde{\Phi}_e \tilde{\cal Q}_e\right],
\end{eqnarray}
where now $\tilde{\cal{M}}$ represents the mass of the non-spinning charged configuration, which from the transformation (\ref{boost}) acquires a new structure:
\textcolor{black}{\begin{eqnarray}\label{masschangecharged}
{\cal{M}} &=& \left[1+\frac{(z+d_{\text{\tiny{eff}}}) w^2}{d_{\text{\tiny{eff}}} (1-w^2)}\right] \, \tilde{\cal{M}}+\left(\frac{d_{\tiny{\mbox{eff}}}}{z+d_{\tiny{\mbox{eff}}}}-\alpha_e\right)\frac{(z+d_{\tiny{\mbox{eff}}}) w^2}{d_{\tiny{\mbox{eff}}} (1-w^2)}\tilde{\Phi}_e\tilde{\cal Q}_e,
\end{eqnarray}
and} finally, we can complete the relation between the entropy and the temperature, given previously in eq. (\ref{eq:S-T}), in the following way
\textcolor{black}{\begin{eqnarray*}
{\cal S}&=&\frac{2 \pi(z+d_{\mbox{\tiny{eff}}})
}{d_{\mbox{\tiny{eff}}}}
\left[-\frac{d_{\mbox{\tiny{eff}}}}{z} \left({\cal{M}}_{\mbox{\tiny{sol}}}+
\Omega_{\mbox{\tiny{sol}}}\,{\cal{J}}_{\mbox{\tiny{sol}}}\right)-\alpha_{m}\, \Phi_m {\cal Q}_m \right] T^{\frac{d_{\mbox{\tiny{eff}}}}{z}} .
\end{eqnarray*}
With} all the above, by using the Cardy-like formula  (\ref{Cardyrotatinghvm}) and starting with the mass of the soliton (uncharged or charged configuration) denoted as $\tilde{\cal{M}}_{\tiny{\mbox{sol}}}$, we have that
its angular momentum becomes
 \begin{eqnarray}
{\cal{J}}_{\tiny{\mbox{sol}}}=-\frac{\tilde{\cal{M}}_{\tiny{\mbox{sol}}}}{z}\, \frac{w (z+d_{\tiny{\mbox{eff}}})}{\left(1-\omega^{2}\right)^{{\frac{3z+d_{\mbox{\tiny{eff}}}}{2z}}}},
\end{eqnarray}
and its mass transforms as
\begin{eqnarray}\label{msolchange}
{\cal{M}}_{\tiny{\mbox{sol}}}= \frac{(z+
d_{\mbox{\tiny{eff}}}\omega^{2})\tilde{\cal{M}}_{\tiny{\mbox{sol}}}}
{z\left(1-\omega^{2}\right)
^{{\frac{3z+d_{\mbox{\tiny{eff}}}}{2z}}}}.
\end{eqnarray}

In all the thermodynamic parameters of the rotating configurations discussed in the sections \ref{isotropic}-\ref{anisotropicsection}, the computations of the Wald entropy, Hawking temperature, and quasilocal formalism are consistent with the expressions given in \textcolor{black}{Eqs. (\ref{entropychange})–(\ref{msolchange}). Additionally, we verified the formula (\ref{Cardyrotatinghvm}) also in a three-dimensional rotating BH with scalar hair that does not come from the
coordinate transformation (\ref{boost}).}

Finally, it would be interesting to extend the exploration to more concrete examples in arbitrary dimensions, involving new charged rotating configurations beyond purely electric or magnetic charges, \textcolor{black}{as discussed in \cite{Bravo-Gaete:2025vyd}}, thereby providing further support for the spinning Cardy-like formula (\ref{Cardyrotatinghvm}), or potentially leading to its refinement.

\acknowledgments
\textcolor{black}{The authors would like to thank to the anonymous referee for carefully reading our manuscript and giving valuable suggestions that led to an improved version of this
work.} MB is supported by Proyecto Interno  UCM-IN-25202 l\'inea regular. XZ is supported by National Natural Science Foundation of China (NSFC) with Grants No.12275087.


\end{document}